\documentclass[sigconf]{acmart}
\usepackage{graphicx}
\usepackage{float}
\usepackage{subfig}
\usepackage{enumitem}
\usepackage{booktabs}
\usepackage{makecell}
\usepackage{arydshln}
\usepackage{marvosym}
\AtBeginDocument{%
  }

\setcopyright{acmlicensed}
\copyrightyear{2025}
\acmYear{2025}
\acmDOI{XXXXXXX.XXXXXXX}
\acmConference[ACM MM]{Make sure to enter the correct
  conference title from your rights confirmation email}{2025}{Dublin, Ireland}
\acmISBN{978-1-4503-XXXX-X/2025/10}
\begin{document}

\title{Deconfounded Reasoning for Multimodal Fake News Detection via Causal Intervention}

\author{
Moyang Liu\textsuperscript{\textmd{1}}, Kaiying Yan\textsuperscript{\textmd{2}}, 
Yukun Liu\textsuperscript{\textmd{3}\Letter}, Ruibo Fu\textsuperscript{\textmd{4}}, 
Zhengqi Wen\textsuperscript{\textmd{5}}, Xuefei Liu\textsuperscript{\textmd{4}},
Chenxing Li\textsuperscript{\textmd{6}}
}

\email{{moyang_liu}@buaa.edu.cn, {yanky}@mail2.sysu.edu.cn, {yukunliu927}@gmail.com}
\affiliation{
\institution{\textsuperscript{\textmd{1}}Beihang University, Beijing, China}
\institution{\textsuperscript{\textmd{2}}School of Mathematics, Sun Yat-sen University, Guangzhou, China}
\institution{\textsuperscript{\textmd{3}}School of Artificial Intelligence, University of Chinese Academy of Sciences, Beijing, China}
\institution{\textsuperscript{\textmd{4}}Institute of Automation, Chinese Academy of Sciences, Beijing, China}
\institution{\textsuperscript{\textmd{5}}Beijing National Research Center for Information Science and Technology, Tsinghua University, Beijing, China}
\institution{\textsuperscript{\textmd{6}}Tencent AI Lab}
\country{}
}
   
\thanks{\Letter\ Corresponding author.}

\renewcommand{\authors}{Moyang Liu, Kaiying Yan, Yukun Liu, Ruibo Fu, Zhengqi Wen, Xuefei Liu, Chenxing Li}
\renewcommand{\shortauthors}{Moyang Liu et al.}


\begin{abstract}
  The rapid growth of social media has led to the widespread dissemination of fake news across multiple content forms, including text, images, audio, and video. Traditional unimodal detection methods fall short in addressing complex cross-modal manipulations; as a result, multimodal fake news detection has emerged as a more effective solution. However, existing multimodal approaches, especially in the context of fake news detection on social media, often overlook the confounders hidden within complex cross-modal interactions, leading models to rely on spurious statistical correlations rather than genuine causal mechanisms. In this paper, we propose the Causal Intervention-based Multimodal Deconfounded Detection (CIMDD) framework, which systematically models three types of confounders via a unified Structural Causal Model (SCM): (1) Lexical Semantic Confounder (LSC); (2) Latent Visual Confounder (LVC); (3) Dynamic Cross-Modal Coupling Confounder (DCCC). To mitigate the influence of these confounders, we specifically design three causal modules based on backdoor adjustment, frontdoor adjustment, and cross-modal joint intervention to block spurious correlations from different perspectives and achieve causal disentanglement of representations for deconfounded reasoning. Experimental results on the FakeSV and FVC datasets demonstrate that CIMDD significantly improves detection accuracy, outperforming state-of-the-art methods by 4.27\% and 4.80\%, respectively. Furthermore, extensive experimental results indicate that CIMDD exhibits strong generalization and robustness across diverse multimodal scenarios.
\end{abstract}



\received{20 February 2007}
\received[revised]{12 March 2009}
\received[accepted]{5 June 2009}

\maketitle

\section{Introduction}

With the rapid growth of social media, fake news has proliferated across platforms, increasingly appearing in multimodal forms such as text, images, audio, and video\cite{capuano2023content}. This diversity makes detection more challenging, as users often spread misinformation without verification\cite{shu2017fake}, leading to widespread misunderstanding and even social unrest\cite{d2021fake}.

Traditional unimodal detection approaches, such as textual semantic analysis\cite{oshikawa2018survey} or image authenticity verification\cite{giachanou2020multimodal}, fail to fully leverage the comprehensive information across all modalities and struggle to capture logical inconsistencies arising from cross-modal coordinated deception\cite{liu2024exploring}. 
Multimodal fake news detection has emerged in this context, aiming to comprehensively analyze information from multiple modalities\cite{tufchi2023comprehensive}. Existing approaches predominantly focus on multimodal information fusion\cite{liu2024exploring}, leveraging cross-attention mechanisms to model interactions between different modalities\cite{khattar2019mvae}. In addition, some methods prioritize textual or visual cues as the primary sources of evidence while incorporating information from other modalities to enhance detection performance\cite{alonso2021sentiment,zhang2023multimodal}. Moreover, given the current scarcity of datasets with full modalities, Qi et al.\cite{qi2023fakesv} and Han et al.\cite{hu2021fvc} have respectively constructed the FakeSV dataset and the FVC dataset, including various contents such as titles, videos, audios, and comments.

\begin{figure*}[!t]
\centering
\subfloat[Lexical Semantic Confounder (LSC)]{\includegraphics[width=1.2in]{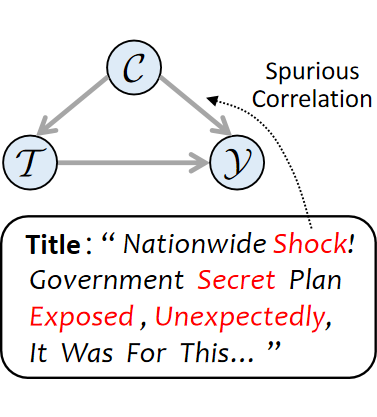}%
\
\label{fig_first_case}}
\hfil
\subfloat[Latent Visual Confounder (LVC)]{\includegraphics[width=2.8in]{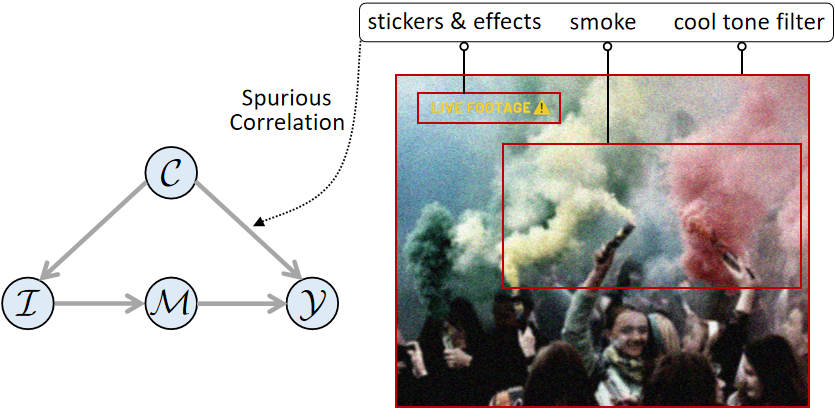}%
\label{fig_second_case}}
\hfil
\subfloat[Dynamic Cross-Modal Coupling Confounder (DCCC)]{\includegraphics[width=2.95in]{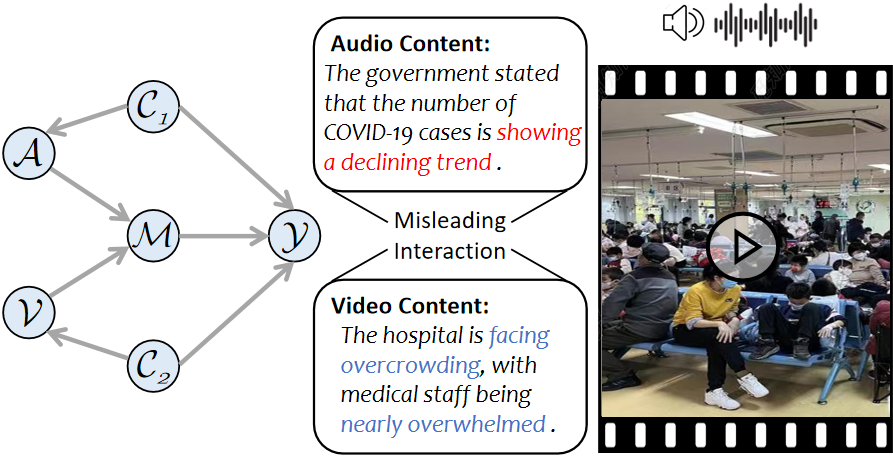}%
\label{fig_third_case}}
\caption{The proposed three types of confounders (LSC, LVC, DCCC) in the context of multimodal fake news detection on social media.}
\label{f1}
\end{figure*}

However, in the task of detecting fake news on social media, various confounders exist that can affect model performance. These confounders become even more complex under the interactions of multimodal data, amplifying their detrimental impact on model reasoning. As a result, these interfering factors establish spurious associations with content veracity, severely impeding models from capturing the underlying causal mechanisms, which in turn undermines their capacity to make accurate predictions. For instance, certain headlines employ emotionally polarizing vocabulary to obscure meaning; images exploit specific filters to induce visual cognitive biases; and audio-visual modalities rely on misleading editing that disrupts the temporal consistency across modalities\cite{gao2024causal}. As a consequence, existing detection models tend to misinterpret these superficial statistical correlations as definitive evidence of false content, thereby failing to capture the genuine causal relationships underpinning the news, which leads to inaccurate detection. Consequently, disentangling confounders from authentic causal features within multimodal representations has emerged as a core challenge in enhancing model robustness and generalizability\cite{yao2021survey}.

To overcome the cognitive limitations imposed by superficial correlations and to identify genuine causal effects\cite{pearl2016causal}, this study establishes a unified structural causal model (SCM)\cite{pearl2009causal} to represent multimodal fake news detection in social media contexts, incorporating three types of confounders arising from interactions among different modalities : (1)Lexical Semantic Confounder (LSC): Spurious word correlations mislead models to rely on superficial lexical cues over genuine content. (2)Latent Visual Confounder (LVC): Latent visual cues, like style or composition, spuriously correlate with veracity and mislead models despite lacking causal relevance. (3)Dynamic Cross-Modal Coupling Confounder (DCCC): Non-causal interactions between modalities (e.g., audio-visual mismatches) mislead models by inducing spurious signals across modalities. These confounders obstruct the model’s ability to learn genuine causal effects from multimodal feature interactions, ultimately leading to degraded detection performance\cite{feder2022causal}.

To mitigate the influence of confounders, we propose the Causal Intervention-based Multimodal Deconfounded Detection (CIMDD) framework grounded in causal inference theory. This framework leverages intervention techniques to block the impact of confounders on fake news detection from a multi-view perspective, thereby enabling causal decision-making.
First, we establish a unified structural causal model (SCM) to represent multimodal fake news detection in social media contexts, incorporating the three confounders mentioned above: LSC, LVC, and DCCC. Then, leveraging the do-calculus in causal intervention, we design three causal modules tailored to each confounder type, namely Linguistic Backdoor Deconfounded Reasoning (LBDR), Visual Frontdoor Deconfounded Reasoning (VFDR), and Cross-modal Joint Deconfounded Reasoning (CJDR). Specifically, we employ backdoor adjustment, frontdoor adjustment, and cross-modal joint intervention to remove confounding effects at different levels, effectively decoupling confounders from multimodal representations. This systematic intervention mechanism facilitates independent learning of multimodal causal representations, ensuring that the model performs deconfounded reasoning.
Extensive experimental results demonstrate that the CIMDD framework effectively identifies causal relationships, significantly improving the accuracy of multimodal fake news detection. Moreover, experiments conducted on diverse datasets reveal the critical role of the proposed causal intervention modules in multimodal learning, highlighting their strong generalization ability and robustness.

The main contributions of this work are summarized as follows:

\begin{itemize}[leftmargin=18pt]
\item We propose a unified structural causal model for multimodal fake news detection in social media contexts and explicitly model three types of confounders arising from cross-modal interactions.
\item We propose the CIMDD framework, which incorporates three causal modules: LBDR, VFDR, and CJDR. These modules are designed using do-operations based on causal intervention techniques, enabling deconfounded reasoning.
\item Extensive experiments on different datasets demonstrate that CIMDD and its modules can significantly improve the performance of multimodal fake news detection while exhibiting strong generalization and robustness.
\end{itemize}

\section{Related Works}

\subsection{Multimodal Fake News Detection}

Existing research on multimodal fake news detection primarily revolves around two paradigms: clue-based analysis and multimodal fusion modeling\cite{comito2023multimodal}.

Clue-based approaches detect fake news by mining specific clues from different modalities (such as text, images, videos, social context, etc.) and analyzing the consistency between features across modalities\cite{xue2021detecting}. Zhu et al.\cite{zhu2022generalizing} treat named entities in news as key clues and dynamically debias these entity clues from a causal perspective. Qi et al.\cite{qi2021improving} explicitly capture three types of cross-modal clues and enhance detection reliability by leveraging consistency among features across modalities. Nan et al.\cite{nan2025exploiting} mine multimodal clues from historical comments and inject them into a student model that relies solely on news content via knowledge distillation. Hu et al.\cite{hu2024bad} leverage large language models to generate multidimensional analysis clues (description, commonsense, factuality) and design modules to evaluate the usefulness of these clues.

Multimodal Fusion Approaches focus on how to effectively integrate information from different modalities, enhancing fake news detection performance through fusion strategies\cite{yan2025mtpareto}. Jing et al.\cite{jing2023multimodal} proposed a progressive fusion network (MPFN) for multimodal disinformation detection, which captures the representational information of each modality at different levels. Choi et al.\cite{choi2021using} constructed FANVN, incorporating adversarial neural networks to extract topic-independent features and using a dynamic gating mechanism for multimodal fusion. Qi et al.\cite{qi2023fakesv} developed the SV-FEND framework, which strengthens the interaction between news content features through a cross-modal attention mechanism and integrates social context information, such as publisher profiles and comments, for comprehensive judgment. Ren et al.\cite{ren2024mmsfd} proposed a multi-grained fusion network (MMSFD), which constructs a multimodal representation that balances local details and global semantics.

However, existing methods are susceptible to confounding factors during multimodal data interaction, leading to reduced generalization \cite{hu2022causal}. In this study, we examine the characteristics of multimodal data in the context of social media from the perspective of causality and disentangle multimodal features and confounding factors through causal intervention techniques.

\subsection{Causal Inference}

As a classical technique in statistics, causal inference has been increasingly integrated with deep learning methods in recent years to eliminate spurious correlations and enhance model generalization \cite{wu2023mfir}. Counterfactual reasoning and causal intervention, two key techniques in causal inference, have been applied across various domains such as Visual Question Answering (VQA) and image captioning\cite{zeng2023correcting}. Niu et al. \cite{niu2021counterfactual} significantly reduced language bias in VQA models by leveraging counterfactual reasoning and causal effect decomposition. Yang et al.\cite{yang2021deconfounded} applied causal intervention in the image captioning task to block the influence of dataset biases. Zeng et al.\cite{zeng2024mitigating} proposed a multi-perspective debiasing framework, MMVD, which learns unbiased multimodal dependencies through counterfactual reasoning.

These approaches demonstrate the effectiveness of integrating causal inference techniques into deep learning models to address various types of biases and improve robustness across different tasks\cite{wang2024vision}. However, existing methods lack a comprehensive confounder hypothesis for multimodal scenarios and struggle to eliminate cross-modal confounders. In this study, we propose a unified structural causal model for multimodal fake news detection, which incorporates three categories of confounders. By leveraging causal intervention techniques, our approach successfully mitigates the influence of these confounders arising from cross-modal interactions within the multimodal features.

\section{Stuctural Causal Model}

The Structural Causal Model (SCM) is a mathematical framework used in causal inference to define causal relationships between variables. It employs directed acyclic graphs (DAGs) to represent causal systems, where nodes denote variables and directed edges (→) represent causal relations between the variables. For instance, $X$ → $Y$ denotes that $X$ is the cause of the effect $Y$.

\begin{figure}[t]
    \centering
    \setlength{\abovecaptionskip}{-0cm}
    \includegraphics[width=0.41\textwidth]{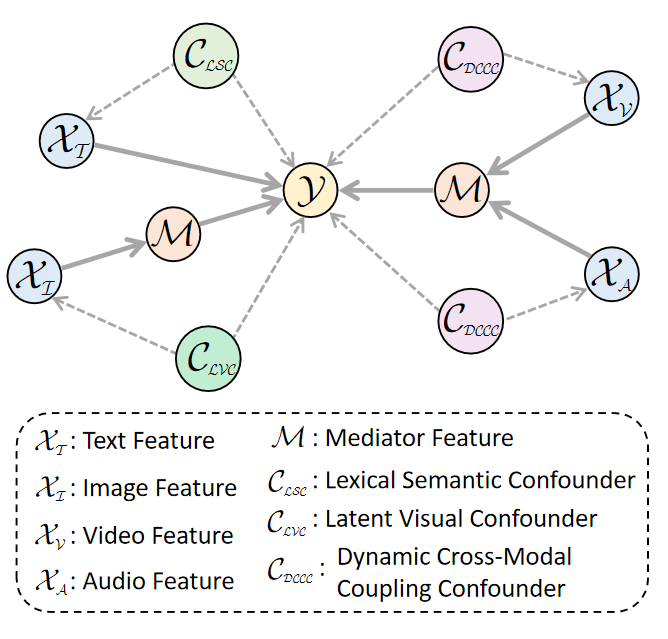}
    \caption{Illustration of the structural causal model for multimodal fake news detection in the context of social media.}
    \label{f2}
\end{figure}

\subsection{SCM in Multimodal Fake News Detection}

As illustrated in Figure \ref{f2}, we construct a unified structural causal model for the task of multimodal fake news detection in social media contexts. It incorporates the three types of confounders C defined above, which are connected via directed edges to the key variables: the news veracity label $Y$ and the input modality features $X=\left\{X_{T}, X_{I}, X_{T}, X_{A}\right\}$. Here, $X_{T}, X_{I}, X_{V}, X_{A}$ represent the textual, image, video, and audio modalities, respectively. These confounders induce spurious correlations between $X$ and $Y$ through backdoor paths of the form $X$ ← $C$ → $Y$. This SCM framework allows us to characterize the spurious associations induced by confounders and guides the design of intervention strategies to recover the true causal effect of multimodal features on news veracity.

\begin{figure*}[ht]
    \centering
    \setlength{\abovecaptionskip}{-0cm}
    \includegraphics[width=17cm]{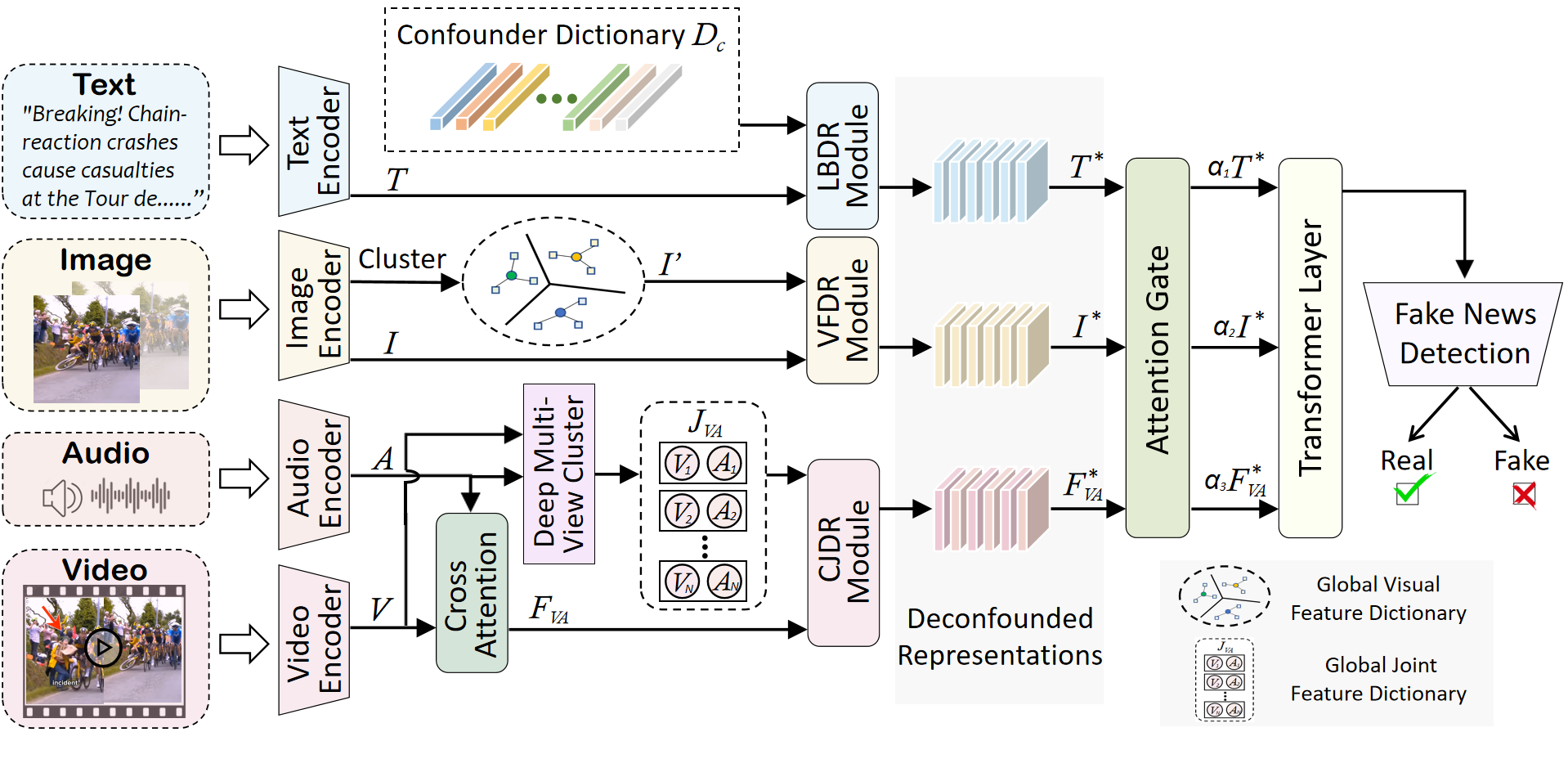}
    \caption{Architecture of the Causal Intervention-based Multimodal Deconfounded Detection framework CIMDD.}
    \label{f3}
\end{figure*}

\subsection{Explicit Modeling of Multimodal Confounders}

To apply the proposed SCM framework, we explicitly model three distinct types of confounders that arise in multimodal fake news detection. These confounders include Lexical Semantic Confounder (LSC), Latent Visual Confounder (LVC), and Dynamic Cross-Modal Coupling Confounder (DCCC), each capturing a different source of spurious associations within and across modalities. The following provides a detailed definition of each confounder:

(1) Lexical Semantic Confounder (LSC): In the textual modality, there exists a plethora of words that are highly correlated with fake news yet lack a causal relationship—such as emotional terms and high-frequency words associated with specific topics. This phenomenon leads models to over-rely on these surface-level features rather than the underlying substantive content. For example, user-generated posts frequently contain emotionally polarizing vocabulary (e.g., “stunning scandal”) and pseudo-logical sentence structures (e.g., “all experts believe that…”), which may mislead the model and result in misclassifications.

(2) Latent Visual Confounder (LVC): In the image modality, numerous latent visual features are unrelated to the veracity of news but can inadvertently mislead models in their authenticity assessments due to implicit correlations. One example of LVC is the interference caused by visual style, such as the application of filters or compositional rules. Unlike the Latent Semantic Confounder (LSC) in the textual modality, these confounders are unobservable and difficult to quantify using conventional statistical methods.

(3) Dynamic Cross-Modal Coupling Confounder (DCCC): In the context of social media, misleading interactions often arise between video and audio modalities due to temporal synchronization or content association (e.g., discrepancies between background music and scene transitions or mismatches between speech tone and visual sentiment). For instance, some videos may deliberately employ solemn background music combined with rapid scene changes to evoke a sense of urgency, even when the content itself is not deceptive. This non-causal cross-modal coupling can mislead models into classifying such content as fake news. Unlike LSC and LVC, which stem from intra-modal feature biases, DCCC emerges from inter-modal interactions, making it particularly challenging to detect through traditional feature engineering or statistical analysis.

\section{Framework of CIMDD}

\subsection{Overview}

The overview of the framework CIMDD is shown in Figure \ref{f3}. This framework incorporates three causal intervention modules: Linguistic Backdoor Deconfounded Reasoning (LBDR), Visual Frontdoor Deconfounded Reasoning (VFDR), and Cross-modal Joint Deconfounded Reasoning (CJDR), which respectively block the effects of the confounders LSC, LVC, and DCCC on the input features. Specifically, the input of each modality is first processed by a modality-specific encoder and then passed into the corresponding causal intervention module to obtain deconfounded multimodal representations. Subsequently, the deconfounded features are passed through an attention-based gating\cite{zhang2020attention} mechanism to obtain their corresponding weights. These features are then weighted accordingly and fed into a transformer\cite{vaswani2017attention} layer. The output of this layer is finally classified into real or fake categories by a classifier.

\begin{figure}[t]
    \centering
    \setlength{\abovecaptionskip}{-0cm}
    \includegraphics[width=0.4\textwidth]{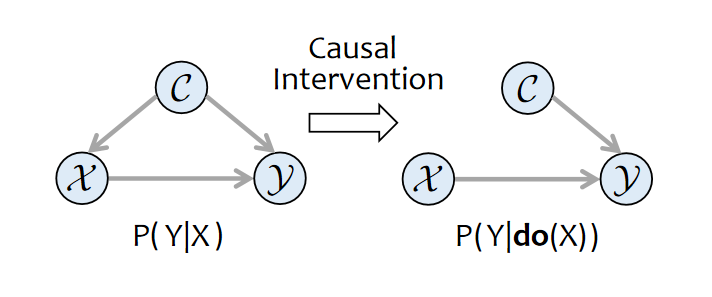}
    \caption{Illustration of causal intervention to block backdoor paths and eliminate confounding bias.}
    \label{f4}
\end{figure}

Before detailing the implementation of each causal module, we first outline the theoretical foundation of causal intervention\cite{pearl2018book}.
Causal intervention techniques aim to eliminate spurious associations introduced by confounders and uncover the true causal effect from observational data. Traditional supervised learning is based on the conditional distribution $P(Y|X)$, which implicitly incorporates the influence of confounders through backdoor paths. In contrast, causal intervention employs the do-operator to estimate $P(Y|do(X))$, which emulates an idealized randomized experiment where $X$ is externally manipulated while all backdoor paths are blocked. 
In SCM, the intervening operation on a variable, as shown in Figure \ref{f4}, removes all edges pointing to it, such that its parent nodes no longer cause it. This operation effectively blocks the paths from $C$ to $X$, ensuring that the model infers outcomes based solely on the direct causal effect of $X$ on $Y$.

\subsection{Linguistic Backdoor Deconfounded Reasoning}

The Linguistic Backdoor Deconfounded Reasoning (LBDR) module applies backdoor adjustment to eliminate the spurious correlations introduced by the Lexical Semantic Confounder (LSC), which is observable in the text modality. Formally, based on the Bayes Rule, the conventional likelihood can be re-written as:

\vspace{-10pt}
\begin{equation}
P(Y|X) = \sum_{c} P(Y,c|X) = \sum_{c} P(Y|X,c) \underline{P(c|X)}
\end{equation}

where $P(c|X)$ reflects spurious associations between $X$ and $Y$ introduced by the effect of confounder $C$. By using the \textit{do-operator}, we can remove the influence of the confounder $C$ on $X$, leading to the following backdoor adjustment formulation: 

\vspace{-10pt}
\begin{align}
P(Y|d o(X)) & =\sum_{c} P(Y, c|d o(X)) \\
& =\sum_{c} P(Y|d o(X), c) P(c|d o(X)) \\
& =\sum_{c} P(Y|X, c) \underline{P(c)} \label{e4}
\end{align}

To implement the backdoor adjustment, we explicitly model the Lexical Semantic Confounder (LSC). Specifically, we construct a lexicon of deceptive-language indicators $\mathbf{L}_{c} = [{l}_{1}, {l}_{2}, \ldots, {l}_{N}] \in \mathbb{R}^{N}$ based on the LIWC2022 dictionary\cite{boyd2022development}, summarizing 20 categories of high-frequency lexical patterns commonly observed in fake news. Based on this lexicon, we can compute the prior probability distribution as follows:

\vspace{-10pt}
\begin{equation}
P(c) = \frac{|l_{\kappa(c)}|}{\sum_{j=1}^{N} |l_j|}, \quad \forall c \in \mathbf{L}_c
\end{equation}

where $\kappa(z)$ denotes the category index of word $z$, and $|l_j|$ denotes the number of occurrences of lexical category $l_{i}$ in the training set. In this way, Equation (\ref{e4}) can be rewritten as: 

\vspace{-10pt}
\begin{align}
P(Y|d o(X)) & =\mathbb{E}_{c}[P(Y|X, c)] \\
& =\mathbb{E}_{c}[\sigma(F(\mathbf{x}, \mathbf{c}))]
\end{align}

where $F(\cdot)$ is a feature fusion model and $\sigma(\cdot)$ denotes the sigmoid function. However, our objective is not to directly estimate the final prediction probability, but rather to obtain a deconfounded feature representation that can be effectively integrated with other modalities to enable deconfounded multimodal prediction. To achieve this, we implement the Normalized Weighted Geometric Mean (NWGM)\cite{xu2015show}, which approximates the expectation over confounder-conditioned features and produces a deconfounded representation:

\vspace{-10pt}
\begin{equation}
    P(Y|d o(X))=\mathbb{E}_{c}[\sigma(F(\mathbf{x}, \mathbf{c}))] \stackrel{N W G M}{\approx} \sigma\left(\mathbb{E}_{c}[F(\mathbf{x}, \mathbf{c})]\right) .
\end{equation}

\begin{figure}[t]
    \centering
    \setlength{\abovecaptionskip}{-0cm}
    \includegraphics[width=0.5\textwidth]{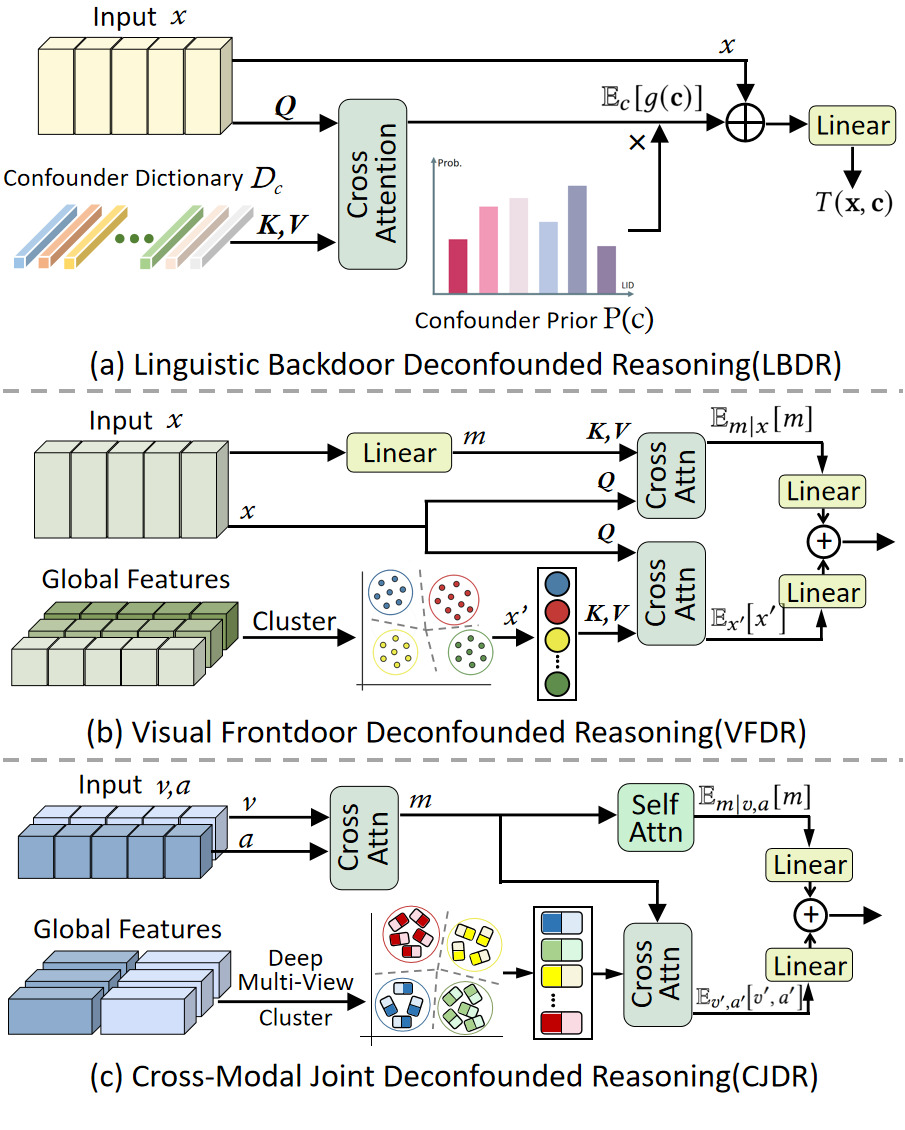}
    \caption{The specific deconfounding processes of the three causal modules (LBDR, VFDR, CJDR).}
    \label{f5}
    \vspace{-0.6cm}
\end{figure}

In this way, we obtain the deconfounded representation of the textual modality: $T(\mathbf{x}, \mathbf{c})=\mathbb{E}_{c}[F(\mathbf{x}, \mathbf{c})])$. Based on the lexicon, we construct a confounder dictionary $\mathbf{D}_{c} = [\mathbf{c}_{1}, \mathbf{c}_{2}, \ldots, \mathbf{c}_{N}] \in \mathbb{R}^{N \times d_c}$, where $N$ denotes the number of word categories and $d_c$ is the dimensionality of the hidden feature space. For each textual input $x$, we compute a confounder count vector $\mathbf{C} = [\mathbf{n}_{1}, \mathbf{n}_{2}, \ldots, \mathbf{n}_{N}] \in \mathbb{R}^{N}$ where $n_i$ represents the number of times words from the 
i-th confounder category appears in the input. We then obtain a sample-specific confounder representation by performing a weighted combination over the dictionary:

\vspace{-10pt}
\begin{equation}
\mathbf{D}_{c}^{*}=\mathbf{C} \odot \mathbf{D}_{c}
\end{equation}

To model the feature fusion function $F(\mathbf{x}, \mathbf{c})$, we adopt a feed-forward layer of the form:

\vspace{-10pt}
\begin{equation}
F(\mathbf{x}, \mathbf{c})=W[\mathbf{x} ; g(\mathbf{c})]    
\end{equation}

where $W$ is a learnable weight matrix, $[;]$ denotes concatenation, and $g(c)$ is a transformation that maps the confounder representation into a feature space of a higher level. Under this formulation, the expected value becomes:

\vspace{-10pt}
\begin{equation}
\mathbb{E}_{\mathbf{c}}[F(\mathbf{x}, \mathbf{c})]=W\left[\mathbf{x} ; \mathbb{E}_{\mathbf{c}}[g(\mathbf{c})]\right]
\end{equation}

Inspired by previous works\cite{chen2023causal,tian2022debiasing}, we implement $g(c)$ using a scaled dot-product attention mechanism over the confounder dictionary $\mathbf{D}_{c}^{*}$. Specifically, we compute:

\vspace{-5pt}
\begin{equation}
\mathbb{E}_{c}[g(\mathbf{c})]=\sum_{u}\left[\operatorname{softmax}\left(\frac{\mathbf{Q}^{T} \mathbf{K}}{\sqrt{d_{m}}}\right) \odot \mathbf{D}_{c}^{*}\right] P(\mathbf{c}),
\end{equation}

where $Q=W_{q}x$ and $K =W_k\mathbf{D}_{c}^{*}$ are the query and key projections, $W_q$ and $Wk$ are learnable parameters, $d_m$ is a scaling factor.

\subsection{Visual Frontdoor Deconfounded Reasoning}

For the Latent Visual Confounder (LVC), its unobservable nature makes it infeasible to be explicitly identified and modeled in advance. For these unobservable confounders, we adopt the front-door adjustment strategy to mitigate its influence, which operates by leveraging a mediator variable M to transmit the causal effect.

In general, attention-based models select specific knowledge representations $M$ from the input features $X$ to predict the output $Y$, thereby forming a predictive path $X$ → $M$ → $Y$\cite{yang2021causal}. This process can be represented as:

\vspace{-10pt}
\begin{equation}
    P(Y|X)=\sum_{m} P(Y|m) P(m|X)
\end{equation}

In multimodal fake news detection on social media, models rely on visual clues such as objects, scenes, and visual composition to infer the veracity of the news. Therefore, we select visual semantics as the mediator variable $M$ in our front-door adjustment framework. By applying the \textit{do-operator} on both the input features $X$ and the mediator variable $M$, we can eliminate the spurious associations introduced by the unobservable confounder LVC, leading to the following front-door adjustment formulation:

\vspace{-10pt}
\begin{align}
P\left(Y | do(X)\right) & =\sum_{m} P\left(Y | do(m)\right) P\left(m | do(X)\right) \\
& =\sum_{m} P(m | X) \sum_{x^{\prime}} P\left(Y | x^{\prime}, m\right) P(x^{\prime})
\end{align}

where $x^{\prime}$ denotes the full set of potential input representations in the visual modality, and the theoretical derivation of the formula will be detailed in the Supplementary Material. $x^{\prime}$ is modelled as a global feature dictionary that approximates the input representation space. Specifically, it is constructed by applying K-means clustering to the visual features of all training samples and randomly sampling one representative instance from each cluster. Similarly, through the NWGM approximation, we obtain the deconfounded representation of the visual modality:

\vspace{-10pt}
\begin{align}
I\left(x^{\prime}, m\right) & \stackrel{N W G M}{\approx}\sum_{m} P(m | X) \sum_{x^{\prime}}G\left(x^{\prime}, m\right) P(x^{\prime}) \\
&\quad = \quad \mathbb{E}_{x^{\prime}} \mathbb{E}_{m \mid x}\left[G\left(x^{\prime}, m\right)\right]
\end{align}

Specifically, we use a linear layer to model $G(\mathbf{x^{\prime}}, \mathbf{m})=\mathbf{W}_{x} \mathbf{x^{\prime}} + \mathbf{W}_{m} \mathbf{m}$, where ${W}_{x}$ and ${W}_{m}$ are learnable weight parameters. In this case, the deconfounded representation $I\left(x^{\prime}, m\right)$ can be simplified as follows:

\vspace{-15pt}
\begin{align}
I\left(x^{\prime}, m\right) &  =  {W}_{x}\cdot\mathbb{E}_{x^{\prime}} [x^{\prime}] + {W}_{m}\cdot\mathbb{E}_{m \mid x} [m]
\end{align}

When computing $\mathbb{E}_{x^{\prime}} [x^{\prime}]$ and $\mathbb{E}_{m \mid x} [m]$, the high dimensionality and sparsity of the sample space make it infeasible to estimate their prior distributions using traditional statistical methods. Therefore, we adopt an attention-based approach to approximate the underlying distributions over both $x^{\prime}$ and $m$, which are estimated via a query-based mechanism. Specifically, two embedding functions $q_{1}(\cdot)$ and $q_{2}(\cdot)$ are used to generate the query vectors $Q_{x_{1}}=q_{1}(x)$ and $Q_{x_{2}}=q_{2}(x)$ from input $x$. Using scaled dot-product attention, the front-door adjustment can be approximated as follows:

\vspace{-10pt}
\begin{align}
\mathbb{E}_{m \mid x}[m] & =\text { Attention }\left(Q_{x_{1}}, K_{m}, V_{m}\right) \nonumber \\
& =\sum_{i} \operatorname{softmax}\left(\frac{Q_{x_{1}} K_{m}^{T}}{\sqrt{d}}\right)_{i} \cdot m_{i} \\
\mathbb{E}_{x^{\prime}}\left[x^{\prime}\right] & =\text { Attention }\left(Q_{x_{2}}, K_{x^{\prime}}, V_{x^{\prime}}\right) \nonumber \\
& =\sum_{k} \operatorname{softmax}\left(\frac{Q_{x_{2}} K_{x^{\prime}}^{T}}{\sqrt{d}}\right)_{k} \cdot x_{k}^{\prime}
\end{align}

where $K_m = V_m = \{m_i\}$ denotes the visual semantic representations, $K_{x^{\prime}} = V_{x^{\prime}} = \{x^{\prime}_k\}$ corresponds to the global visual feature dictionary and $d$ is the dimensionality of the key vectors used for scaling. Thus, we have obtained the deconfounded representation of the visual modality through complete front-door adjustment.

\begin{table*}[ht]
\caption{Performance comparison of CIMDD and other models on the FakeSV and FVC datasets.}
\begin{tabular}{lcccccccc}
\Xhline{1.2pt}
\multicolumn{1}{c}{} & \multicolumn{4}{c}{FakeSV}                                        & \multicolumn{4}{c}{FVC}                                                           \\ \hline
Method               & Accuracy       & F1-score       & Precision      & Recall         & Accuracy       & F1-score               & Precision              & Recall         \\ \hline
VGGish+SVM(Audio)\cite{hershey2017cnn}    & 61.25          & 61.31          & 61.24          & 61.33          & 58.44          & 58.61                  & 58.48                  & 58.63          \\
VGG19+Att(Image)\cite{simonyan2014very}     & 68.53          & 68.51          & 68.53          & 68.50          & 65.79          & 65.81                  & 65.49                  & 66.08          \\
C3D+Att(Video)\cite{tran2015learning}       & 70.26          & 70.24          & 70.25          & 70.25          & 71.81          & 71.72                  & 71.89                  & 71.85          \\
Bert+Att(Text)\cite{devlin2019bert}       & 74.31          & 74.35          & 74.30          & 74.39          & 76.37          & 76.35                  & 76.39                  & 76.33          \\ \hline
FANVN\textit{(2021)}\cite{choi2021using}                & 75.04          & 75.02          & 75.11          & 75.04          & 85.81          & 85.32                  & 85.20                  & 85.44          \\
SV-FEND\textit{(2023)}\cite{qi2023fakesv}              & 79.31          & 79.24          & 79.62          & 79.31          & 84.71          & 85.37                  & 84.25                  & 86.53          \\
SVRPM\textit{(2024)} \cite{wu2024interpretable}               & 79.34          & 78.55          & 79.75          & 78.16          & -              & -                      & -                      & -              \\ 
MMSFD\textit{(2024)}\cite{ren2024mmsfd}                & 81.83          & 81.81          & 82.02          & 81.81          & -              & -                       & -                 
    & -              \\ 
MMVD\textit{(2024)}\cite{zeng2024mitigating}                 & 82.64          & 82.63          & 82.63          & 82.73          & 89.28          & 90.36                  & 90.27                  & 90.46          \\ \hline
\textbf{CIMDD}       & \textbf{86.91} & \textbf{84.82} & \textbf{85.53} & \textbf{84.47} & \textbf{94.08} & \textbf{93.27}         & \textbf{94.15}         & \textbf{92.74} \\ \Xhline{1.2pt}
\end{tabular}
\label{t1}
\end{table*}

\subsection{Cross-Modal Joint Deconfounded Reasoning}

Unlike LSC and LVC, which are confined to individual modalities, DCCC arises from semantic and temporal interactions across modalities. As it only appears during multimodal integration, existing deconfounding techniques like backdoor and frontdoor adjustment, which are primarily designed for unimodal inputs, are insufficient to eliminate spurious correlations during multimodal fusion.

To address this limitation, we innovatively propose a cross-modal joint intervention mechanism grounded in causal theory. Due to the inherent temporal characteristics of video and audio modalities, most existing models employ cross-attention mechanisms to fuse video and audio features. Given the input video features $V$ and audio features $A$, the cross-modal fused representation $M$ is computed as follows:

\vspace{-9pt}
\begin{equation}
\begin{gathered}
Q = W_Q \cdot V, \quad K = W_K \cdot A, \quad V' = W_V \cdot A \\
M = \mathrm{softmax}\left( \frac{Q \cdot K^T}{\sqrt{d}\,} \right) \cdot V'
\end{gathered}
\end{equation}

The model then uses the fused features for prediction. This process can be presented as:

\vspace{-5pt}
\begin{equation}
    P(Y|V,A)=\sum_{m} P(Y|m) P(m|V,A)
\end{equation}

To eliminate the spurious correlations introduced by cross-modal interactions, we apply the \textit{do-operator} on video features $V$, audio features $A$, and the fused representation $M$, leading to the following cross-modal joint intervention formulation:

\vspace{-10pt}
\begin{align}
P\left(Y | do(V,A)\right) & =\sum_{m} P\left(Y | do(m)\right) P\left(m | do(V,A)\right) \\
& =\sum_{m} P(m | V,A) \sum_{v^{\prime},a^{\prime}} P\left(Y | v^{\prime}, a^{\prime}, m\right) P(v^{\prime},a^{\prime})
\end{align}

The detailed theoretical proof and derivation of this formulation are provided in the Supplementary Material. Similar to the approach used in Section 4.3, we apply the NWGM approximation along with a linear mapping model to approximate the expected intervention effect. Accordingly, the cross-modal deconfounded representation can be obtained as follows:

\vspace{-10pt}
\begin{align}
R\left(v^{\prime},a^{\prime}, m\right) & \stackrel{N W G M}{\approx}\sum_{m} P(m | V,A) \sum_{v^{\prime},a^{\prime}}g\left(v^{\prime},a^{\prime}, m\right) P(v^{\prime},a^{\prime}) \\
&\quad = \quad \mathbb{E}_{v^{\prime},a^{\prime}} \mathbb{E}_{m \mid v,a}\left[g\left(v^{\prime},a^{\prime}, m\right)\right]\\
&\quad  =  \quad {W}_{ab}\cdot\mathbb{E}_{v^{\prime},a^{\prime}} [v^{\prime},a^{\prime}] + {W}_{m}\cdot\mathbb{E}_{m \mid v,a} [m]
\end{align}

Motivated by the Law of Large Numbers, we approximate the expectation over the true joint distribution $P(v^{\prime},a^{\prime})$ using the empirical distribution over observed training samples $(v^{\prime}_i, a^{\prime}_i)$:

\vspace{-10pt}
\begin{equation}
    \mathbb{E}_{v^{\prime}, a^{\prime}}\left[v^{\prime}, a^{\prime}\right] \approx \mathbb{E}_{\left(v^{\prime}, a^{\prime}\right) \sim \hat{P}_{\mathrm{emp}}}[v^{\prime}, a^{\prime}]
\end{equation}

In this case, we perform multi-view clustering on the video and audio features $v_i$ and $a_i$ using subspace learning and contrastive learning, and randomly sample one representative instance from each cluster to construct a global joint feature dictionary $J_{VA}=\left[\left(v_{1}, a_{1}\right),\left(v_{2}, a_{2}\right), \ldots,\left(v_{N}, a_{N}\right)\right]$. The detailed procedure of multi-view clustering is provided in the Supplementary Material.

For the estimation of $\mathbb{E}_{m \mid v,a}[m]$ and $\mathbb{E}_{v^{\prime}, a^{\prime}}\left[v^{\prime}, a^{\prime}\right]$, we still follow a similar approach to that in Section 4.3. Through a fusion function $f(\cdot)$ and embedding functions $q(\cdot)$, the query matrix $Q$ is obtained from the input features $V$ and $A$ as $Q_{f_1}=q_1(f(V,A))$ and $Q_{f_2}=q_2(f(V,A))$. The expectations can be approximated as follows:

\vspace{-10pt}
\begin{align}
\mathbb{E}_{m \mid v,a} [m] & =\text { Attention }\left(Q_{f_{1}}, K_{m}, V_{m}\right) \nonumber \\
& =\sum_{i} \operatorname{softmax}\left(\frac{Q_{f_{1}} K_{m}^{T}}{\sqrt{d}}\right)_{i} \cdot m_{i} \\
\mathbb{E}_{v^{\prime}, a^{\prime}}\left[v^{\prime}, a^{\prime}\right] & =\text { Attention }\left(Q_{f_{2}}, K_{va}, V_{va}\right) \nonumber \\
& =\sum_{k} \operatorname{softmax}\left(\frac{Q_{f_{2}} K_{va}^{T}}{\sqrt{d}}\right)_{k} \cdot f(v_{k}^{\prime},a_{k}^{\prime})
\end{align}

where $K_m = V_m = \{m_i\}$ denotes the dynamic fusion features, $K_{va} = V_{va} = f(v_{k}^{\prime},a_{k}^{\prime})$ corresponds to the global fusion feature dictionary. In this study, we implement the fusion function $f(\cdot)$ using a cross-attention mechanism.

\section{Experiments}

\subsection{Datasets and Implementation Details}

\subsubsection{Datasets}

We used two multimodal datasets in the context of social media, FakeSV and FVC. The details of the datasets are described as follows:

\begin{itemize}[leftmargin=18pt]
\item The FakeSV dataset, constructed by Qi et al.\cite{qi2023fakesv}, consists of an extensive collection of Chinese news short videos. It incorporates multiple modalities, including text, video, audio, and social context, enabling the comprehensive representation of diverse social media scenarios.
\item The FVC dataset\cite{hu2021fvc} comprises genuine and fake videos sourced from YouTube, covering various topics such as politics, sports, and entertainment. It integrates multiple modalities, including video titles, video content, user comments, and URLs, making it particularly valuable for research in fake news detection.
\end{itemize}

\subsubsection{Implementation Details}
In the experiments, the dataset was divided into training, validation, and test sets in a 70:15:15 ratio following a chronological order. The model utilized the cross-entropy loss function and AdamW\cite{loshchilov2017decoupled} optimizer, with a batch size of 64 and a learning rate of 5e-5. All experiments were conducted on NVIDIA RTX 4090 GPUs. The evaluation metrics used to assess model performance were accuracy, F1-score, Precision, and Recall. The final results were obtained by evaluating this best model on the test set. In the feature extraction stage, we employed modality-specific models: BERT for textual modality, Wav2Vec2.0\cite{baevski2020wav2vec} for audio modality, CLIP\cite{radford2021learning} for image modality, and Timesformer\cite{bertasius2021space} for video clip modality. For the LBDR module, the number of categories ($N$) in the confounder dictionary was set to 20 (details of these categories are provided in the Supplementary Material), and the dimension of hidden features ($d_c$) was configured as 64. In both the VFDR and CJDR modules, we set the number of categories in the global visual feature dictionary and the joint feature dictionary to 128.

\begin{table}[t]
\caption{Performance comparison of baseline models with and without Causal Intervention(CI) methods.}
\begin{tabular}{lcccc}
\Xhline{1.2pt}
                        & \multicolumn{2}{c}{FakeSV}      & \multicolumn{2}{c}{FVC}         \\ \hline
Method                  & Accuracy       & F1-score       & Accuracy       & F1-score       \\ \hline
\textbf{FANVN}                   & 75.04          & 75.02          & 85.81          & 85.32          \\
\multicolumn{1}{c}{+CI} & \textbf{78.60} & \textbf{78.46} & \textbf{86.35} & \textbf{86.25} \\ \hdashline[2pt/5pt]
\textbf{SV-FEND}                 & 79.31          & 79.24          & 84.71          & 85.37          \\
\multicolumn{1}{c}{+CI} & \textbf{82.29} & \textbf{82.02} & \textbf{86.47} & \textbf{86.39} \\ \Xhline{1.2pt}
\end{tabular}
\label{t2}
\vspace{-10pt}
\end{table}

\subsection{Experimental Results}

Table \ref{t1} and Table \ref{t2} present the experimental performance of our proposed framework CIMDD on two benchmark datasets. An initial observation from Table \ref{t1} is that multimodal fake news detection methods generally outperform unimodal ones, which explains why most recent approaches are designed to leverage information from multiple modalities. This integration enables models to better capture complementary and diverse cues from text, audio, image, and video, contributing to improved detection performance. 

Building upon these insights, CIMDD further advances multimodal fake news detection by explicitly addressing key limitations in existing approaches. As shown in Table \ref{t1}, CIMDD consistently achieves the best performance across all evaluation metrics among the recent and comprehensive detection models. Specifically, CIMDD outperforms the state-of-the-art methods by 4.27\% and 4.80\% in terms of accuracy on the FakeSV and FVC datasets, respectively, highlighting its excellent generalization ability and robustness across different scenarios. These results demonstrate that CIMDD effectively eliminates the influence of confounders, accurately identifies genuine causal effects, and successfully achieves deconfounded reasoning.

Table \ref{t2} presents the experimental results of applying our proposed systematic causal intervention methodology to several baseline models. These results were obtained by integrating the three causal intervention modules(LBDR, VFDR, and CJDR) into the publicly available implementations of the respective baselines. As shown in Table \ref{t2}, incorporating our causal intervention framework leads to performance improvements across most evaluation metrics when compared to the original multimodal fake news detection models.
Specifically, the accuracy of baseline models FANVN\cite{choi2021using} and SV-FEND\cite{qi2023fakesv} on the FakeSV dataset improved by 3.56\% and 2.98\%, respectively, after integrating our method. Similarly, on the FVC dataset, the accuracy gains were 0.54\% and 1.76\% for the same models. These results demonstrate that our proposed framework, CIMDD, effectively mitigates the influence of confounders through causal interventions and exhibits strong generalization ability and robustness across different datasets.

\subsection{Ablation Studies}

To evaluate the effectiveness of the proposed causal intervention strategy and each individual module within the CIMDD framework, we conducted two sets of experiments. Table \ref{t3} presents the results obtained by removing specific modules from the full CIMDD framework to evaluate their individual contributions. In contrast, Table \ref{t4} shows the performance gains obtained by integrating the corresponding causal modules into unimodal baseline models, enabling evaluation of their benefits within each individual modality.

The results in Table \ref{t4} demonstrate that incorporating the corresponding causal intervention modules into unimodal models consistently improves performance across modalities, further supporting the general applicability and effectiveness of each module.

\begin{table}[t]
\caption{Impact of Causal Intervention(CI) and causal modules(LBDR, VFDR, CJDR) in CIMDD}
\begin{tabular}{lcccc}
\Xhline{1.2pt}
Method         & Accuracy       & F1-score       & Precision      & Recall         \\ \hline
w/o CI         & 81.73          & 81.23          & 81.84          & 80.93          \\
w/o LBDR        & 83.03          & 82.67          & 82.92          & 82.50          \\
w/o VFDR        & 84.13          & 83.81          & 84.02          & 83.67          \\
w/o CJDR       & 85.06          & 84.58          & 85.52          & 84.17          \\ \hline
\textbf{CIMDD} & \textbf{86.91} & \textbf{84.82} & \textbf{85.53} & \textbf{84.47} \\ \Xhline{1.2pt}
\end{tabular}
\label{t3}
\end{table}

\begin{table}[t]
\caption{Impact of causal modules on unimodal baselines}
\begin{tabular}{lcccc}
\Xhline{1.2pt}
Method                    & Acc.       & F1       & Prec.      & Recall         \\ \hline
\textbf{Bert+Att(Text)}            & 74.31          & 74.35          & 74.30          & 74.39          \\
\multicolumn{1}{c}{+LBDR}  & \textbf{77.49} & \textbf{77.06} & \textbf{77.19} & \textbf{76.97} \\ \hdashline[2pt/5pt]
\textbf{VGG19+Att(Image)}          & 68.53          & 68.51          & 68.53          & 68.50          \\
\multicolumn{1}{c}{+VFDR}  & \textbf{71.07} & \textbf{71.32} & \textbf{71.68} & \textbf{71.12} \\ \hdashline[2pt/5pt]
\textbf{Fusion(Video+Audio)}       & 73.99          & 73.45          & 73.62          & 73.34          \\
\multicolumn{1}{c}{+CJDR} & \textbf{75.54} & \textbf{75.03} & \textbf{74.79} & \textbf{75.52} \\ \Xhline{1.2pt}
\end{tabular}
\label{t4}
\vspace{-0.2cm}
\end{table}

The results in Table \ref{t3} clearly show that removing any of the causal modules leads to a noticeable drop in performance, which strongly validates the necessity and effectiveness of the proposed causal intervention components. Notably, the complete removal of causal intervention (CIMDD w/o CI) results in the most substantial performance degradation across all metrics, further emphasizing the overall importance of causal intervention in reducing confounding bias and enhancing both the robustness and generalization capability of the model.

Among the three modules, Linguistic Backdoor Deconfounded Reasoning (LBDR) has the most significant impact on overall performance. This suggests that lexical semantic confounders introduce the greatest amount of spurious correlations, misleading the model during training. This observation is consistent with the nature of social media content, where emotionally charged or logically misleading phrases frequently appear and correlate spuriously with fake news labels.

The second most impactful component is Visual Frontdoor Deconfounded Reasoning (VFDR), indicating that latent visual confounders also considerably hinder the model’s ability to capture true causal signals. Many hidden visual cues may correlate with fake news labels but are not causally informative, thus impairing model generalization if not properly handled.

Lastly, Cross-modal Joint Deconfounded Reasoning (CJDR) contributes to performance improvement by mitigating the influence of Dynamic Cross-Modal Coupling Confounders, which can arise from complex and misleading interactions between modalities.

\begin{figure}[t]
    \centering
    \setlength{\abovecaptionskip}{-0cm}
    \includegraphics[width=0.5\textwidth]{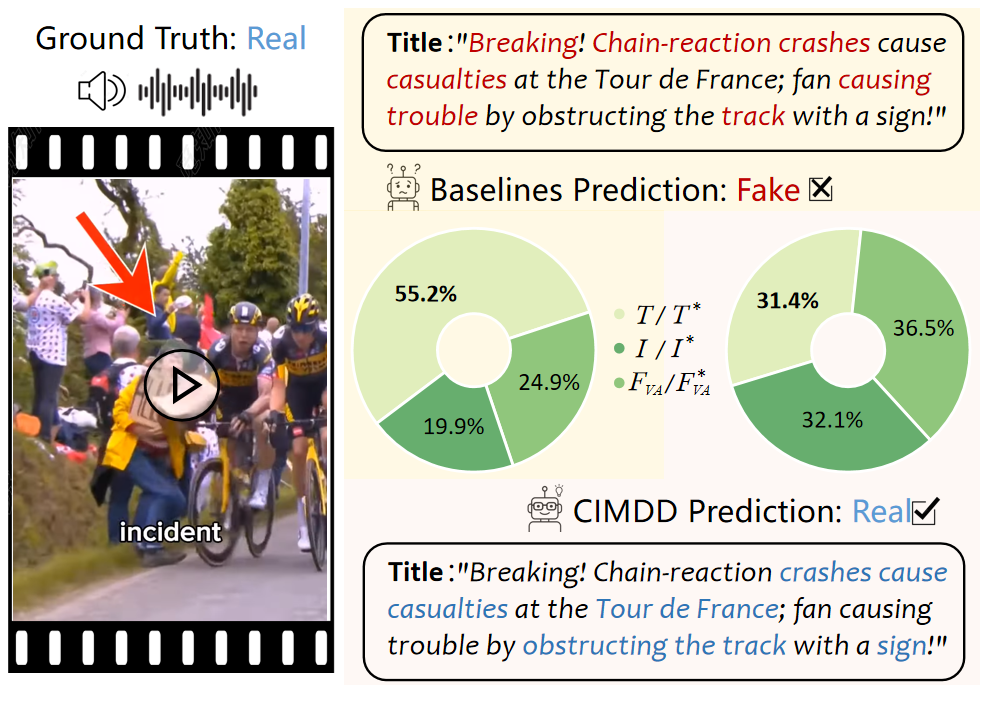}
    \caption{A specific news case from the FakeSV dataset.}
    \label{f6}
    \vspace{-0.3cm}
\end{figure}

\subsection{Case Study}

To better understand how the causal intervention mechanism contributes to fake news detection, we conduct a case study focusing on the textual modality. Specifically, we compare the predictions, attention distributions from the text encoder, and modality-level feature weights of CIMDD and a baseline model on a representative example. As shown in Figure \ref{f6}, the baseline model places excessive attention on lexical cues that are highly correlated with fake news and prone to misinterpretation, such as “Breaking” and “casualties.” This not only amplifies the influence of these confounding words within the textual representation but also leads to an overestimation of the text modality’s importance during fusion, ultimately resulting in a misclassification as fake news. In contrast, CIMDD effectively mitigates the impact of lexical semantic confounders by redirecting attention toward more causally relevant and informative content and successfully identifies the input as real news through the Linguistic Backdoor Deconfounded Reasoning module.

\section{Conclusion}

In this paper, we propose a unified Structural Causal Model (SCM) for multimodal fake news detection in social media contexts. Within the SCM framework, we explicitly model and analyze three types of confounders that commonly arise in cross-modal feature interactions. To eliminate spurious correlations introduced by these confounders, we develop a causal intervention-based framework called CIMDD. Specifically, we apply backdoor adjustment, frontdoor adjustment, and cross-modal joint intervention to design three corresponding causal modules: LBDR, VFDR, and CJDR, each of which targets a specific type of confounder. Extensive experiments and ablation studies on the FakeSV and FVC datasets demonstrate that CIMDD and its individual components significantly improve detection accuracy while also exhibiting strong generalization ability and robustness across diverse scenarios.

\bibliographystyle{ACM-Reference-Format}
\bibliography{main}


\begin{thebibliography}{52}


\ifx \showCODEN    \undefined \def \showCODEN     #1{\unskip}     \fi
\ifx \showISBNx    \undefined \def \showISBNx     #1{\unskip}     \fi
\ifx \showISBNxiii \undefined \def \showISBNxiii  #1{\unskip}     \fi
\ifx \showISSN     \undefined \def \showISSN      #1{\unskip}     \fi
\ifx \showLCCN     \undefined \def \showLCCN      #1{\unskip}     \fi
\ifx \shownote     \undefined \def \shownote      #1{#1}          \fi
\ifx \showarticletitle \undefined \def \showarticletitle #1{#1}   \fi
\ifx \showURL      \undefined \def \showURL       {\relax}        \fi
\providecommand\bibfield[2]{#2}
\providecommand\bibinfo[2]{#2}
\providecommand\natexlab[1]{#1}
\providecommand\showeprint[2][]{arXiv:#2}

\bibitem[Alonso et~al\mbox{.}(2021)]%
        {alonso2021sentiment}
\bibfield{author}{\bibinfo{person}{Miguel~A Alonso}, \bibinfo{person}{David Vilares}, \bibinfo{person}{Carlos G{\'o}mez-Rodr{\'\i}guez}, {and} \bibinfo{person}{Jes{\'u}s Vilares}.} \bibinfo{year}{2021}\natexlab{}.
\newblock \showarticletitle{Sentiment analysis for fake news detection}.
\newblock \bibinfo{journal}{\emph{Electronics}} \bibinfo{volume}{10}, \bibinfo{number}{11} (\bibinfo{year}{2021}), \bibinfo{pages}{1348}.
\newblock


\bibitem[Baevski et~al\mbox{.}(2020)]%
        {baevski2020wav2vec}
\bibfield{author}{\bibinfo{person}{Alexei Baevski}, \bibinfo{person}{Yuhao Zhou}, \bibinfo{person}{Abdelrahman Mohamed}, {and} \bibinfo{person}{Michael Auli}.} \bibinfo{year}{2020}\natexlab{}.
\newblock \showarticletitle{wav2vec 2.0: A framework for self-supervised learning of speech representations}.
\newblock \bibinfo{journal}{\emph{Advances in neural information processing systems}}  \bibinfo{volume}{33} (\bibinfo{year}{2020}), \bibinfo{pages}{12449--12460}.
\newblock


\bibitem[Bertasius et~al\mbox{.}(2021)]%
        {bertasius2021space}
\bibfield{author}{\bibinfo{person}{Gedas Bertasius}, \bibinfo{person}{Heng Wang}, {and} \bibinfo{person}{Lorenzo Torresani}.} \bibinfo{year}{2021}\natexlab{}.
\newblock \showarticletitle{Is space-time attention all you need for video understanding?}. In \bibinfo{booktitle}{\emph{ICML}}, Vol.~\bibinfo{volume}{2}. \bibinfo{pages}{4}.
\newblock


\bibitem[Boyd et~al\mbox{.}(2022)]%
        {boyd2022development}
\bibfield{author}{\bibinfo{person}{Ryan~L Boyd}, \bibinfo{person}{Ashwini Ashokkumar}, \bibinfo{person}{Sarah Seraj}, {and} \bibinfo{person}{James~W Pennebaker}.} \bibinfo{year}{2022}\natexlab{}.
\newblock \showarticletitle{The development and psychometric properties of LIWC-22}.
\newblock \bibinfo{journal}{\emph{Austin, TX: University of Texas at Austin}}  \bibinfo{volume}{10} (\bibinfo{year}{2022}), \bibinfo{pages}{1--47}.
\newblock


\bibitem[Capuano et~al\mbox{.}(2023)]%
        {capuano2023content}
\bibfield{author}{\bibinfo{person}{Nicola Capuano}, \bibinfo{person}{Giuseppe Fenza}, \bibinfo{person}{Vincenzo Loia}, {and} \bibinfo{person}{Francesco~David Nota}.} \bibinfo{year}{2023}\natexlab{}.
\newblock \showarticletitle{Content-based fake news detection with machine and deep learning: A systematic review}.
\newblock \bibinfo{journal}{\emph{Neurocomputing}}  \bibinfo{volume}{530} (\bibinfo{year}{2023}), \bibinfo{pages}{91--103}.
\newblock


\bibitem[Chen et~al\mbox{.}(2023)]%
        {chen2023causal}
\bibfield{author}{\bibinfo{person}{Ziwei Chen}, \bibinfo{person}{Linmei Hu}, \bibinfo{person}{Weixin Li}, \bibinfo{person}{Yingxia Shao}, {and} \bibinfo{person}{Liqiang Nie}.} \bibinfo{year}{2023}\natexlab{}.
\newblock \showarticletitle{Causal intervention and counterfactual reasoning for multi-modal fake news detection}. In \bibinfo{booktitle}{\emph{Proceedings of the 61st Annual Meeting of the Association for Computational Linguistics (Volume 1: Long Papers)}}. \bibinfo{pages}{627--638}.
\newblock


\bibitem[Choi and Ko(2021)]%
        {choi2021using}
\bibfield{author}{\bibinfo{person}{Hyewon Choi} {and} \bibinfo{person}{Youngjoong Ko}.} \bibinfo{year}{2021}\natexlab{}.
\newblock \showarticletitle{Using topic modeling and adversarial neural networks for fake news video detection}. In \bibinfo{booktitle}{\emph{Proceedings of the 30th ACM international conference on information \& knowledge management}}. \bibinfo{pages}{2950--2954}.
\newblock


\bibitem[Comito et~al\mbox{.}(2023)]%
        {comito2023multimodal}
\bibfield{author}{\bibinfo{person}{Carmela Comito}, \bibinfo{person}{Luciano Caroprese}, {and} \bibinfo{person}{Ester Zumpano}.} \bibinfo{year}{2023}\natexlab{}.
\newblock \showarticletitle{Multimodal fake news detection on social media: a survey of deep learning techniques}.
\newblock \bibinfo{journal}{\emph{Social Network Analysis and Mining}} \bibinfo{volume}{13}, \bibinfo{number}{1} (\bibinfo{year}{2023}), \bibinfo{pages}{101}.
\newblock


\bibitem[Devlin et~al\mbox{.}(2019)]%
        {devlin2019bert}
\bibfield{author}{\bibinfo{person}{Jacob Devlin}, \bibinfo{person}{Ming-Wei Chang}, \bibinfo{person}{Kenton Lee}, {and} \bibinfo{person}{Kristina Toutanova}.} \bibinfo{year}{2019}\natexlab{}.
\newblock \showarticletitle{Bert: Pre-training of deep bidirectional transformers for language understanding}. In \bibinfo{booktitle}{\emph{Proceedings of the 2019 conference of the North American chapter of the association for computational linguistics: human language technologies, volume 1 (long and short papers)}}. \bibinfo{pages}{4171--4186}.
\newblock


\bibitem[D’ulizia et~al\mbox{.}(2021)]%
        {d2021fake}
\bibfield{author}{\bibinfo{person}{Arianna D’ulizia}, \bibinfo{person}{Maria~Chiara Caschera}, \bibinfo{person}{Fernando Ferri}, {and} \bibinfo{person}{Patrizia Grifoni}.} \bibinfo{year}{2021}\natexlab{}.
\newblock \showarticletitle{Fake news detection: a survey of evaluation datasets}.
\newblock \bibinfo{journal}{\emph{PeerJ Computer Science}}  \bibinfo{volume}{7} (\bibinfo{year}{2021}), \bibinfo{pages}{e518}.
\newblock


\bibitem[Feder et~al\mbox{.}(2022)]%
        {feder2022causal}
\bibfield{author}{\bibinfo{person}{Amir Feder}, \bibinfo{person}{Katherine~A Keith}, \bibinfo{person}{Emaad Manzoor}, \bibinfo{person}{Reid Pryzant}, \bibinfo{person}{Dhanya Sridhar}, \bibinfo{person}{Zach Wood-Doughty}, \bibinfo{person}{Jacob Eisenstein}, \bibinfo{person}{Justin Grimmer}, \bibinfo{person}{Roi Reichart}, \bibinfo{person}{Margaret~E Roberts}, {et~al\mbox{.}}} \bibinfo{year}{2022}\natexlab{}.
\newblock \showarticletitle{Causal inference in natural language processing: Estimation, prediction, interpretation and beyond}.
\newblock \bibinfo{journal}{\emph{Transactions of the Association for Computational Linguistics}}  \bibinfo{volume}{10} (\bibinfo{year}{2022}), \bibinfo{pages}{1138--1158}.
\newblock


\bibitem[Gao et~al\mbox{.}(2024)]%
        {gao2024causal}
\bibfield{author}{\bibinfo{person}{Chen Gao}, \bibinfo{person}{Yu Zheng}, \bibinfo{person}{Wenjie Wang}, \bibinfo{person}{Fuli Feng}, \bibinfo{person}{Xiangnan He}, {and} \bibinfo{person}{Yong Li}.} \bibinfo{year}{2024}\natexlab{}.
\newblock \showarticletitle{Causal inference in recommender systems: A survey and future directions}.
\newblock \bibinfo{journal}{\emph{ACM Transactions on Information Systems}} \bibinfo{volume}{42}, \bibinfo{number}{4} (\bibinfo{year}{2024}), \bibinfo{pages}{1--32}.
\newblock


\bibitem[Giachanou et~al\mbox{.}(2020)]%
        {giachanou2020multimodal}
\bibfield{author}{\bibinfo{person}{Anastasia Giachanou}, \bibinfo{person}{Guobiao Zhang}, {and} \bibinfo{person}{Paolo Rosso}.} \bibinfo{year}{2020}\natexlab{}.
\newblock \showarticletitle{Multimodal multi-image fake news detection}. In \bibinfo{booktitle}{\emph{2020 IEEE 7th international conference on data science and advanced analytics (DSAA)}}. IEEE, \bibinfo{pages}{647--654}.
\newblock


\bibitem[Hershey et~al\mbox{.}(2017)]%
        {hershey2017cnn}
\bibfield{author}{\bibinfo{person}{Shawn Hershey}, \bibinfo{person}{Sourish Chaudhuri}, \bibinfo{person}{Daniel~PW Ellis}, \bibinfo{person}{Jort~F Gemmeke}, \bibinfo{person}{Aren Jansen}, \bibinfo{person}{R~Channing Moore}, \bibinfo{person}{Manoj Plakal}, \bibinfo{person}{Devin Platt}, \bibinfo{person}{Rif~A Saurous}, \bibinfo{person}{Bryan Seybold}, {et~al\mbox{.}}} \bibinfo{year}{2017}\natexlab{}.
\newblock \showarticletitle{CNN architectures for large-scale audio classification}. In \bibinfo{booktitle}{\emph{2017 ieee international conference on acoustics, speech and signal processing (icassp)}}. IEEE, \bibinfo{pages}{131--135}.
\newblock


\bibitem[Hu et~al\mbox{.}(2024)]%
        {hu2024bad}
\bibfield{author}{\bibinfo{person}{Beizhe Hu}, \bibinfo{person}{Qiang Sheng}, \bibinfo{person}{Juan Cao}, \bibinfo{person}{Yuhui Shi}, \bibinfo{person}{Yang Li}, \bibinfo{person}{Danding Wang}, {and} \bibinfo{person}{Peng Qi}.} \bibinfo{year}{2024}\natexlab{}.
\newblock \showarticletitle{Bad actor, good advisor: Exploring the role of large language models in fake news detection}. In \bibinfo{booktitle}{\emph{Proceedings of the AAAI Conference on Artificial Intelligence}}, Vol.~\bibinfo{volume}{38}. \bibinfo{pages}{22105--22113}.
\newblock


\bibitem[Hu et~al\mbox{.}(2022)]%
        {hu2022causal}
\bibfield{author}{\bibinfo{person}{Linmei Hu}, \bibinfo{person}{Ziwei Chen}, \bibinfo{person}{Ziwang Zhao}, \bibinfo{person}{Jianhua Yin}, {and} \bibinfo{person}{Liqiang Nie}.} \bibinfo{year}{2022}\natexlab{}.
\newblock \showarticletitle{Causal inference for leveraging image-text matching bias in multi-modal fake news detection}.
\newblock \bibinfo{journal}{\emph{IEEE Transactions on Knowledge and Data Engineering}} \bibinfo{volume}{35}, \bibinfo{number}{11} (\bibinfo{year}{2022}), \bibinfo{pages}{11141--11152}.
\newblock


\bibitem[Hu et~al\mbox{.}(2021)]%
        {hu2021fvc}
\bibfield{author}{\bibinfo{person}{Zhihao Hu}, \bibinfo{person}{Guo Lu}, {and} \bibinfo{person}{Dong Xu}.} \bibinfo{year}{2021}\natexlab{}.
\newblock \showarticletitle{FVC: A new framework towards deep video compression in feature space}. In \bibinfo{booktitle}{\emph{Proceedings of the IEEE/CVF conference on computer vision and pattern recognition}}. \bibinfo{pages}{1502--1511}.
\newblock


\bibitem[Jing et~al\mbox{.}(2023)]%
        {jing2023multimodal}
\bibfield{author}{\bibinfo{person}{Jing Jing}, \bibinfo{person}{Hongchen Wu}, \bibinfo{person}{Jie Sun}, \bibinfo{person}{Xiaochang Fang}, {and} \bibinfo{person}{Huaxiang Zhang}.} \bibinfo{year}{2023}\natexlab{}.
\newblock \showarticletitle{Multimodal fake news detection via progressive fusion networks}.
\newblock \bibinfo{journal}{\emph{Information processing \& management}} \bibinfo{volume}{60}, \bibinfo{number}{1} (\bibinfo{year}{2023}), \bibinfo{pages}{103120}.
\newblock


\bibitem[Khattar et~al\mbox{.}(2019)]%
        {khattar2019mvae}
\bibfield{author}{\bibinfo{person}{Dhruv Khattar}, \bibinfo{person}{Jaipal~Singh Goud}, \bibinfo{person}{Manish Gupta}, {and} \bibinfo{person}{Vasudeva Varma}.} \bibinfo{year}{2019}\natexlab{}.
\newblock \showarticletitle{Mvae: Multimodal variational autoencoder for fake news detection}. In \bibinfo{booktitle}{\emph{The world wide web conference}}. \bibinfo{pages}{2915--2921}.
\newblock


\bibitem[Liu et~al\mbox{.}(2024a)]%
        {liu2024exploring}
\bibfield{author}{\bibinfo{person}{Moyang Liu}, \bibinfo{person}{Yukun Liu}, \bibinfo{person}{Ruibo Fu}, \bibinfo{person}{Zhengqi Wen}, \bibinfo{person}{Jianhua Tao}, \bibinfo{person}{Xuefei Liu}, {and} \bibinfo{person}{Guanjun Li}.} \bibinfo{year}{2024}\natexlab{a}.
\newblock \showarticletitle{Exploring the role of audio in multimodal misinformation detection}. In \bibinfo{booktitle}{\emph{2024 IEEE 14th International Symposium on Chinese Spoken Language Processing (ISCSLP)}}. IEEE, \bibinfo{pages}{204--208}.
\newblock


\bibitem[Liu et~al\mbox{.}(2024b)]%
        {liu2024misd}
\bibfield{author}{\bibinfo{person}{Moyang Liu}, \bibinfo{person}{Kaiying Yan}, \bibinfo{person}{Yukun Liu}, \bibinfo{person}{Ruibo Fu}, \bibinfo{person}{Zhengqi Wen}, \bibinfo{person}{Xuefei Liu}, {and} \bibinfo{person}{Chenxing Li}.} \bibinfo{year}{2024}\natexlab{b}.
\newblock \showarticletitle{MisD-MoE: A Multimodal Misinformation Detection Framework with Adaptive Feature Selection}. In \bibinfo{booktitle}{\emph{NeurIPS Efficient Natural Language and Speech Processing Workshop}}. PMLR, \bibinfo{pages}{114--122}.
\newblock


\bibitem[Loshchilov and Hutter(2017)]%
        {loshchilov2017decoupled}
\bibfield{author}{\bibinfo{person}{Ilya Loshchilov} {and} \bibinfo{person}{Frank Hutter}.} \bibinfo{year}{2017}\natexlab{}.
\newblock \showarticletitle{Decoupled weight decay regularization}.
\newblock \bibinfo{journal}{\emph{arXiv preprint arXiv:1711.05101}} (\bibinfo{year}{2017}).
\newblock


\bibitem[Nan et~al\mbox{.}(2025)]%
        {nan2025exploiting}
\bibfield{author}{\bibinfo{person}{Qiong Nan}, \bibinfo{person}{Qiang Sheng}, \bibinfo{person}{Juan Cao}, \bibinfo{person}{Yongchun Zhu}, \bibinfo{person}{Danding Wang}, \bibinfo{person}{Guang Yang}, {and} \bibinfo{person}{Jintao Li}.} \bibinfo{year}{2025}\natexlab{}.
\newblock \showarticletitle{Exploiting user comments for early detection of fake news prior to users’ commenting}.
\newblock \bibinfo{journal}{\emph{Frontiers of Computer Science}} \bibinfo{volume}{19}, \bibinfo{number}{10} (\bibinfo{year}{2025}), \bibinfo{pages}{1910354}.
\newblock


\bibitem[Niu et~al\mbox{.}(2021)]%
        {niu2021counterfactual}
\bibfield{author}{\bibinfo{person}{Yulei Niu}, \bibinfo{person}{Kaihua Tang}, \bibinfo{person}{Hanwang Zhang}, \bibinfo{person}{Zhiwu Lu}, \bibinfo{person}{Xian-Sheng Hua}, {and} \bibinfo{person}{Ji-Rong Wen}.} \bibinfo{year}{2021}\natexlab{}.
\newblock \showarticletitle{Counterfactual vqa: A cause-effect look at language bias}. In \bibinfo{booktitle}{\emph{Proceedings of the IEEE/CVF conference on computer vision and pattern recognition}}. \bibinfo{pages}{12700--12710}.
\newblock


\bibitem[Oshikawa et~al\mbox{.}(2018)]%
        {oshikawa2018survey}
\bibfield{author}{\bibinfo{person}{Ray Oshikawa}, \bibinfo{person}{Jing Qian}, {and} \bibinfo{person}{William~Yang Wang}.} \bibinfo{year}{2018}\natexlab{}.
\newblock \showarticletitle{A survey on natural language processing for fake news detection}.
\newblock \bibinfo{journal}{\emph{arXiv preprint arXiv:1811.00770}} (\bibinfo{year}{2018}).
\newblock


\bibitem[Pearl(2009)]%
        {pearl2009causal}
\bibfield{author}{\bibinfo{person}{Judea Pearl}.} \bibinfo{year}{2009}\natexlab{}.
\newblock \showarticletitle{Causal inference in statistics: An overview}.
\newblock  (\bibinfo{year}{2009}).
\newblock


\bibitem[Pearl et~al\mbox{.}(2016)]%
        {pearl2016causal}
\bibfield{author}{\bibinfo{person}{Judea Pearl}, \bibinfo{person}{Madelyn Glymour}, {and} \bibinfo{person}{Nicholas~P Jewell}.} \bibinfo{year}{2016}\natexlab{}.
\newblock \bibinfo{booktitle}{\emph{Causal inference in statistics: A primer}}.
\newblock \bibinfo{publisher}{John Wiley \& Sons}.
\newblock


\bibitem[Pearl and Mackenzie(2018)]%
        {pearl2018book}
\bibfield{author}{\bibinfo{person}{Judea Pearl} {and} \bibinfo{person}{Dana Mackenzie}.} \bibinfo{year}{2018}\natexlab{}.
\newblock \bibinfo{booktitle}{\emph{The book of why: the new science of cause and effect}}.
\newblock \bibinfo{publisher}{Basic books}.
\newblock


\bibitem[Qi et~al\mbox{.}(2023)]%
        {qi2023fakesv}
\bibfield{author}{\bibinfo{person}{Peng Qi}, \bibinfo{person}{Yuyan Bu}, \bibinfo{person}{Juan Cao}, \bibinfo{person}{Wei Ji}, \bibinfo{person}{Ruihao Shui}, \bibinfo{person}{Junbin Xiao}, \bibinfo{person}{Danding Wang}, {and} \bibinfo{person}{Tat-Seng Chua}.} \bibinfo{year}{2023}\natexlab{}.
\newblock \showarticletitle{Fakesv: A multimodal benchmark with rich social context for fake news detection on short video platforms}. In \bibinfo{booktitle}{\emph{Proceedings of the AAAI Conference on Artificial Intelligence}}, Vol.~\bibinfo{volume}{37}. \bibinfo{pages}{14444--14452}.
\newblock


\bibitem[Qi et~al\mbox{.}(2021)]%
        {qi2021improving}
\bibfield{author}{\bibinfo{person}{Peng Qi}, \bibinfo{person}{Juan Cao}, \bibinfo{person}{Xirong Li}, \bibinfo{person}{Huan Liu}, \bibinfo{person}{Qiang Sheng}, \bibinfo{person}{Xiaoyue Mi}, \bibinfo{person}{Qin He}, \bibinfo{person}{Yongbiao Lv}, \bibinfo{person}{Chenyang Guo}, {and} \bibinfo{person}{Yingchao Yu}.} \bibinfo{year}{2021}\natexlab{}.
\newblock \showarticletitle{Improving fake news detection by using an entity-enhanced framework to fuse diverse multimodal clues}. In \bibinfo{booktitle}{\emph{Proceedings of the 29th ACM International Conference on Multimedia}}. \bibinfo{pages}{1212--1220}.
\newblock


\bibitem[Radford et~al\mbox{.}(2021)]%
        {radford2021learning}
\bibfield{author}{\bibinfo{person}{Alec Radford}, \bibinfo{person}{Jong~Wook Kim}, \bibinfo{person}{Chris Hallacy}, \bibinfo{person}{Aditya Ramesh}, \bibinfo{person}{Gabriel Goh}, \bibinfo{person}{Sandhini Agarwal}, \bibinfo{person}{Girish Sastry}, \bibinfo{person}{Amanda Askell}, \bibinfo{person}{Pamela Mishkin}, \bibinfo{person}{Jack Clark}, {et~al\mbox{.}}} \bibinfo{year}{2021}\natexlab{}.
\newblock \showarticletitle{Learning transferable visual models from natural language supervision}. In \bibinfo{booktitle}{\emph{International conference on machine learning}}. PmLR, \bibinfo{pages}{8748--8763}.
\newblock


\bibitem[Ren et~al\mbox{.}(2024)]%
        {ren2024mmsfd}
\bibfield{author}{\bibinfo{person}{Shuai Ren}, \bibinfo{person}{Yahui Liu}, \bibinfo{person}{Yaping Zhu}, \bibinfo{person}{Wanlong Bing}, \bibinfo{person}{Hongliang Ma}, {and} \bibinfo{person}{Wei Wang}.} \bibinfo{year}{2024}\natexlab{}.
\newblock \showarticletitle{MMSFD: Multi-grained and Multi-modal Fusion for Short Video Fake News Detection}. In \bibinfo{booktitle}{\emph{2024 7th International Conference on Data Science and Information Technology (DSIT)}}. IEEE, \bibinfo{pages}{1--11}.
\newblock


\bibitem[Shu et~al\mbox{.}(2017)]%
        {shu2017fake}
\bibfield{author}{\bibinfo{person}{Kai Shu}, \bibinfo{person}{Amy Sliva}, \bibinfo{person}{Suhang Wang}, \bibinfo{person}{Jiliang Tang}, {and} \bibinfo{person}{Huan Liu}.} \bibinfo{year}{2017}\natexlab{}.
\newblock \showarticletitle{Fake news detection on social media: A data mining perspective}.
\newblock \bibinfo{journal}{\emph{ACM SIGKDD explorations newsletter}} \bibinfo{volume}{19}, \bibinfo{number}{1} (\bibinfo{year}{2017}), \bibinfo{pages}{22--36}.
\newblock


\bibitem[Simonyan and Zisserman(2014)]%
        {simonyan2014very}
\bibfield{author}{\bibinfo{person}{Karen Simonyan} {and} \bibinfo{person}{Andrew Zisserman}.} \bibinfo{year}{2014}\natexlab{}.
\newblock \showarticletitle{Very deep convolutional networks for large-scale image recognition}.
\newblock \bibinfo{journal}{\emph{arXiv preprint arXiv:1409.1556}} (\bibinfo{year}{2014}).
\newblock


\bibitem[Tian et~al\mbox{.}(2022)]%
        {tian2022debiasing}
\bibfield{author}{\bibinfo{person}{Bing Tian}, \bibinfo{person}{Yixin Cao}, \bibinfo{person}{Yong Zhang}, {and} \bibinfo{person}{Chunxiao Xing}.} \bibinfo{year}{2022}\natexlab{}.
\newblock \showarticletitle{Debiasing nlu models via causal intervention and counterfactual reasoning}. In \bibinfo{booktitle}{\emph{Proceedings of the AAAI Conference on Artificial Intelligence}}, Vol.~\bibinfo{volume}{36}. \bibinfo{pages}{11376--11384}.
\newblock


\bibitem[Tran et~al\mbox{.}(2015)]%
        {tran2015learning}
\bibfield{author}{\bibinfo{person}{Du Tran}, \bibinfo{person}{Lubomir Bourdev}, \bibinfo{person}{Rob Fergus}, \bibinfo{person}{Lorenzo Torresani}, {and} \bibinfo{person}{Manohar Paluri}.} \bibinfo{year}{2015}\natexlab{}.
\newblock \showarticletitle{Learning spatiotemporal features with 3d convolutional networks}. In \bibinfo{booktitle}{\emph{Proceedings of the IEEE international conference on computer vision}}. \bibinfo{pages}{4489--4497}.
\newblock


\bibitem[Tufchi et~al\mbox{.}(2023)]%
        {tufchi2023comprehensive}
\bibfield{author}{\bibinfo{person}{Shivani Tufchi}, \bibinfo{person}{Ashima Yadav}, {and} \bibinfo{person}{Tanveer Ahmed}.} \bibinfo{year}{2023}\natexlab{}.
\newblock \showarticletitle{A comprehensive survey of multimodal fake news detection techniques: advances, challenges, and opportunities}.
\newblock \bibinfo{journal}{\emph{International Journal of Multimedia Information Retrieval}} \bibinfo{volume}{12}, \bibinfo{number}{2} (\bibinfo{year}{2023}), \bibinfo{pages}{28}.
\newblock


\bibitem[Vaswani et~al\mbox{.}(2017)]%
        {vaswani2017attention}
\bibfield{author}{\bibinfo{person}{Ashish Vaswani}, \bibinfo{person}{Noam Shazeer}, \bibinfo{person}{Niki Parmar}, \bibinfo{person}{Jakob Uszkoreit}, \bibinfo{person}{Llion Jones}, \bibinfo{person}{Aidan~N Gomez}, \bibinfo{person}{{\L}ukasz Kaiser}, {and} \bibinfo{person}{Illia Polosukhin}.} \bibinfo{year}{2017}\natexlab{}.
\newblock \showarticletitle{Attention is all you need}.
\newblock \bibinfo{journal}{\emph{Advances in neural information processing systems}}  \bibinfo{volume}{30} (\bibinfo{year}{2017}).
\newblock


\bibitem[Wang et~al\mbox{.}(2024)]%
        {wang2024vision}
\bibfield{author}{\bibinfo{person}{Liuyi Wang}, \bibinfo{person}{Zongtao He}, \bibinfo{person}{Ronghao Dang}, \bibinfo{person}{Mengjiao Shen}, \bibinfo{person}{Chengju Liu}, {and} \bibinfo{person}{Qijun Chen}.} \bibinfo{year}{2024}\natexlab{}.
\newblock \showarticletitle{Vision-and-language navigation via causal learning}. In \bibinfo{booktitle}{\emph{Proceedings of the IEEE/CVF Conference on Computer Vision and Pattern Recognition}}. \bibinfo{pages}{13139--13150}.
\newblock


\bibitem[Wu et~al\mbox{.}(2024)]%
        {wu2024interpretable}
\bibfield{author}{\bibinfo{person}{Kaixuan Wu}, \bibinfo{person}{Yanghao Lin}, \bibinfo{person}{Donglin Cao}, {and} \bibinfo{person}{Dazhen Lin}.} \bibinfo{year}{2024}\natexlab{}.
\newblock \showarticletitle{Interpretable short video rumor detection based on modality tampering}. In \bibinfo{booktitle}{\emph{Proceedings of the 2024 Joint International Conference on Computational Linguistics, Language Resources and Evaluation (LREC-COLING 2024)}}. \bibinfo{pages}{9180--9189}.
\newblock


\bibitem[Wu et~al\mbox{.}(2023)]%
        {wu2023mfir}
\bibfield{author}{\bibinfo{person}{Lianwei Wu}, \bibinfo{person}{Yuzhou Long}, \bibinfo{person}{Chao Gao}, \bibinfo{person}{Zhen Wang}, {and} \bibinfo{person}{Yanning Zhang}.} \bibinfo{year}{2023}\natexlab{}.
\newblock \showarticletitle{MFIR: Multimodal fusion and inconsistency reasoning for explainable fake news detection}.
\newblock \bibinfo{journal}{\emph{Information Fusion}}  \bibinfo{volume}{100} (\bibinfo{year}{2023}), \bibinfo{pages}{101944}.
\newblock


\bibitem[Xu et~al\mbox{.}(2015)]%
        {xu2015show}
\bibfield{author}{\bibinfo{person}{Kelvin Xu}, \bibinfo{person}{Jimmy Ba}, \bibinfo{person}{Ryan Kiros}, \bibinfo{person}{Kyunghyun Cho}, \bibinfo{person}{Aaron Courville}, \bibinfo{person}{Ruslan Salakhudinov}, \bibinfo{person}{Rich Zemel}, {and} \bibinfo{person}{Yoshua Bengio}.} \bibinfo{year}{2015}\natexlab{}.
\newblock \showarticletitle{Show, attend and tell: Neural image caption generation with visual attention}. In \bibinfo{booktitle}{\emph{International conference on machine learning}}. PMLR, \bibinfo{pages}{2048--2057}.
\newblock


\bibitem[Xue et~al\mbox{.}(2021)]%
        {xue2021detecting}
\bibfield{author}{\bibinfo{person}{Junxiao Xue}, \bibinfo{person}{Yabo Wang}, \bibinfo{person}{Yichen Tian}, \bibinfo{person}{Yafei Li}, \bibinfo{person}{Lei Shi}, {and} \bibinfo{person}{Lin Wei}.} \bibinfo{year}{2021}\natexlab{}.
\newblock \showarticletitle{Detecting fake news by exploring the consistency of multimodal data}.
\newblock \bibinfo{journal}{\emph{Information Processing \& Management}} \bibinfo{volume}{58}, \bibinfo{number}{5} (\bibinfo{year}{2021}), \bibinfo{pages}{102610}.
\newblock


\bibitem[Yan et~al\mbox{.}(2025)]%
        {yan2025mtpareto}
\bibfield{author}{\bibinfo{person}{Kaiying Yan}, \bibinfo{person}{Moyang Liu}, \bibinfo{person}{Yukun Liu}, \bibinfo{person}{Ruibo Fu}, \bibinfo{person}{Zhengqi Wen}, \bibinfo{person}{Jianhua Tao}, \bibinfo{person}{Xuefei Liu}, {and} \bibinfo{person}{Guanjun Li}.} \bibinfo{year}{2025}\natexlab{}.
\newblock \showarticletitle{MTPareto: A MultiModal Targeted Pareto Framework for Fake News Detection}.
\newblock \bibinfo{journal}{\emph{arXiv preprint arXiv:2501.06764}} (\bibinfo{year}{2025}).
\newblock


\bibitem[Yang et~al\mbox{.}(2021a)]%
        {yang2021deconfounded}
\bibfield{author}{\bibinfo{person}{Xu Yang}, \bibinfo{person}{Hanwang Zhang}, {and} \bibinfo{person}{Jianfei Cai}.} \bibinfo{year}{2021}\natexlab{a}.
\newblock \showarticletitle{Deconfounded image captioning: A causal retrospect}.
\newblock \bibinfo{journal}{\emph{IEEE Transactions on Pattern Analysis and Machine Intelligence}} \bibinfo{volume}{45}, \bibinfo{number}{11} (\bibinfo{year}{2021}), \bibinfo{pages}{12996--13010}.
\newblock


\bibitem[Yang et~al\mbox{.}(2021b)]%
        {yang2021causal}
\bibfield{author}{\bibinfo{person}{Xu Yang}, \bibinfo{person}{Hanwang Zhang}, \bibinfo{person}{Guojun Qi}, {and} \bibinfo{person}{Jianfei Cai}.} \bibinfo{year}{2021}\natexlab{b}.
\newblock \showarticletitle{Causal attention for vision-language tasks}. In \bibinfo{booktitle}{\emph{Proceedings of the IEEE/CVF conference on computer vision and pattern recognition}}. \bibinfo{pages}{9847--9857}.
\newblock


\bibitem[Yao et~al\mbox{.}(2021)]%
        {yao2021survey}
\bibfield{author}{\bibinfo{person}{Liuyi Yao}, \bibinfo{person}{Zhixuan Chu}, \bibinfo{person}{Sheng Li}, \bibinfo{person}{Yaliang Li}, \bibinfo{person}{Jing Gao}, {and} \bibinfo{person}{Aidong Zhang}.} \bibinfo{year}{2021}\natexlab{}.
\newblock \showarticletitle{A survey on causal inference}.
\newblock \bibinfo{journal}{\emph{ACM Transactions on Knowledge Discovery from Data (TKDD)}} \bibinfo{volume}{15}, \bibinfo{number}{5} (\bibinfo{year}{2021}), \bibinfo{pages}{1--46}.
\newblock


\bibitem[Zeng et~al\mbox{.}(2024)]%
        {zeng2024mitigating}
\bibfield{author}{\bibinfo{person}{Zhi Zeng}, \bibinfo{person}{Minnan Luo}, \bibinfo{person}{Xiangzheng Kong}, \bibinfo{person}{Huan Liu}, \bibinfo{person}{Hao Guo}, \bibinfo{person}{Hao Yang}, \bibinfo{person}{Zihan Ma}, {and} \bibinfo{person}{Xiang Zhao}.} \bibinfo{year}{2024}\natexlab{}.
\newblock \showarticletitle{Mitigating World Biases: A Multimodal Multi-View Debiasing Framework for Fake News Video Detection}. In \bibinfo{booktitle}{\emph{Proceedings of the 32nd ACM International Conference on Multimedia}}. \bibinfo{pages}{6492--6500}.
\newblock


\bibitem[Zeng et~al\mbox{.}(2023)]%
        {zeng2023correcting}
\bibfield{author}{\bibinfo{person}{Zhi Zeng}, \bibinfo{person}{Mingmin Wu}, \bibinfo{person}{Guodong Li}, \bibinfo{person}{Xiang Li}, \bibinfo{person}{Zhongqiang Huang}, {and} \bibinfo{person}{Ying Sha}.} \bibinfo{year}{2023}\natexlab{}.
\newblock \showarticletitle{Correcting the bias: Mitigating multimodal inconsistency contrastive learning for multimodal fake news detection}. In \bibinfo{booktitle}{\emph{2023 IEEE International Conference on Multimedia and Expo (ICME)}}. IEEE, \bibinfo{pages}{2861--2866}.
\newblock


\bibitem[Zhang et~al\mbox{.}(2020)]%
        {zhang2020attention}
\bibfield{author}{\bibinfo{person}{Jianxin Zhang}, \bibinfo{person}{Zongkang Jiang}, \bibinfo{person}{Jing Dong}, \bibinfo{person}{Yaqing Hou}, {and} \bibinfo{person}{Bin Liu}.} \bibinfo{year}{2020}\natexlab{}.
\newblock \showarticletitle{Attention gate resU-Net for automatic MRI brain tumor segmentation}.
\newblock \bibinfo{journal}{\emph{IEEE Access}}  \bibinfo{volume}{8} (\bibinfo{year}{2020}), \bibinfo{pages}{58533--58545}.
\newblock


\bibitem[Zhang et~al\mbox{.}(2023)]%
        {zhang2023multimodal}
\bibfield{author}{\bibinfo{person}{Xichen Zhang}, \bibinfo{person}{Sajjad Dadkhah}, \bibinfo{person}{Alexander~Gerald Weismann}, \bibinfo{person}{Mohammad~Amin Kanaani}, {and} \bibinfo{person}{Ali~A Ghorbani}.} \bibinfo{year}{2023}\natexlab{}.
\newblock \showarticletitle{Multimodal fake news analysis based on image--text similarity}.
\newblock \bibinfo{journal}{\emph{IEEE Transactions on Computational Social Systems}} \bibinfo{volume}{11}, \bibinfo{number}{1} (\bibinfo{year}{2023}), \bibinfo{pages}{959--972}.
\newblock


\bibitem[Zhu et~al\mbox{.}(2022)]%
        {zhu2022generalizing}
\bibfield{author}{\bibinfo{person}{Yongchun Zhu}, \bibinfo{person}{Qiang Sheng}, \bibinfo{person}{Juan Cao}, \bibinfo{person}{Shuokai Li}, \bibinfo{person}{Danding Wang}, {and} \bibinfo{person}{Fuzhen Zhuang}.} \bibinfo{year}{2022}\natexlab{}.
\newblock \showarticletitle{Generalizing to the future: Mitigating entity bias in fake news detection}. In \bibinfo{booktitle}{\emph{Proceedings of the 45th International ACM SIGIR Conference on Research and Development in Information Retrieval}}. \bibinfo{pages}{2120--2125}.
\newblock


\end{thebibliography}


\end{document}